\documentstyle[12pt]{article}

\begin{document}

\renewcommand{\thefootnote}{\fnsymbol{footnote}}

\begin{titlepage}

\thispagestyle{empty}

\begin{flushright}
KIAS-P01011 \\
hep-th/0102160
\end{flushright}

\vspace{5mm}

\begin{center}
{\Large \bf Noncommutative $Q$-balls}\\[10mm]
Youngjai Kiem${}^{a,}$\footnote{{\tt Email address: ykiem$@$newton.skku.ac.kr}},
Chanju Kim${}^{b,}$\footnote{{\tt Email address: cjkim$@$kias.re.kr}},
and
Yoonbai Kim${}^{a,}$\footnote{{\tt Email address: yoonbai$@$skku.ac.kr}}
\\
{\it ${}^{a}$BK21 Physics Research Division and Institute of Basic Science,
Sungkyunkwan University,\\
Suwon 440-746, Korea}\\
{\it ${}^{b}$School of Physics, Korea Institute for Advanced Study,\\
Seoul 130-012, Korea}
\end{center}

\vspace{5mm}

\begin{abstract}
We obtain $Q$-ball solutions in noncommutative scalar field 
theory with a global U(1) invariance.  The $Q$-ball 
solutions are shown to be classically and quantum 
mechanically stable.  We also find that ``excited $Q$-ball"
states exist for some class of scalar potentials, which are 
classically stable in the large noncommutativity limit.  
\end{abstract}

\vspace{20mm}


\end{titlepage}

\renewcommand{\thefootnote}{\arabic{footnote}}
\setcounter{footnote}{0}

Apart from their string theoretic origin, noncommutative
field theories possess a number of striking features
that cannot be found for commutative 
counterparts \cite{harvey}.
A prime example is the existence of the classically
stable, static soliton solution in noncommutative scalar field
theories in more than (1+1)-dimensions \cite{GMS,various}.  
The infinite number
of derivatives appearing in the $*$-products allows
us to evade the arguments of Derrick and Hobart, which
show that such solitons do not exist.
Historically, one way to evade the arguments within
the framework of ordinary field theories is to
consider the stationary solutions in the presence of
the continuous global symmetry, leading to $Q$-balls
(or nontopological solitons) \cite{FLS,Col,qbb}.
From the viewpoint of physical applications, $Q$-balls 
in supersymmetric theories have recently been suggested
to be a possible candidate of dark matter \cite{kus}.
In this paper, we show that it is possible to fuse these
two lines of approaches.  The analysis of $Q$-balls 
in noncommutative scalar field theories with a global
U(1) invariance can be performed by using the formalism
of Gopakumar, Minwalla and Seiberg (GMS) \cite{GMS}, 
leading to noncommutative $Q$-balls.  

According to the analysis of \cite{GMS}, the natural
vantage point  
to study the GMS solitons is the large noncommutativity 
parameter $\theta$ limit.  In this limit, the spatial
kinetic terms are subleading, and exact nontopological
GMS soliton solutions are obtained upon neglecting them.  
As we turn off the
value of $\theta$, GMS solitons cease to exist.  In the
ordinary $Q$-ball case ($\theta = 0$), the spatial 
kinetic terms can be neglected in the large $Q$ limit
(the thin-wall approximation).  This is the limit
where we have the analytic handle over $Q$-balls.
As we turn on the value
of $\theta$ and crank it up, the $Q$-balls continue to
exist as nontopological solitons resembling the GMS
solitons, as we will show in this paper.  The 
simplification for the $Q$-ball physics
coming from the large $\theta$ limit is that we can relax the
large $Q$ condition and still get the explicit form of the
exact solutions.  

While the noncommutative $Q$-ball solutions
can be obtained by applying the techniques developed for
the construction of GMS solitons, there are important
differences even in the large $\theta$ limit.  While the
classically stable GMS solitons are constructed from the
local minimum points of the potential, some noncommutative
$Q$-balls can have classically stable ``excited
$Q$-ball states", which also involve single local
maximum point of the potential; we will derive the 
condition for the potentials for which such states
exist.  Furthermore, one can show that
the noncommutative $Q$-balls are stable even quantum
mechanically\footnote{Some subtleties related to the 
UV/IR mixing \cite{MRS} that happens in the loop physics of 
noncommutative field theories will be discussed
later in this paper.  We show that our results should 
not be drastically
modified up to one-loop order.}.  
On the other hand, GMS solitons are quantum mechanically 
metastable, since there is no conserved charge that
can protect them from decaying into elementary quanta.   

A simple (commutative) model field theory that possesses $Q$-balls is the
$(2+1)$-dimensional complex scalar theory with a global U(1) 
invariance
\begin{equation}\label{comm}
S=\int dt d^{2}x\left[\frac{1}{2}\partial_{\mu}\bar{\phi} 
 \partial^{\mu}\phi
- V(\bar{\phi},\phi)\right],
\end{equation}
where the scalar potential is given by  
\begin{equation}
 V(\bar{\phi},\phi) = \frac{1}{2}m^{2}\bar{\phi}\phi
+\sum_{k=2}^{N} \frac{b_{k}}{(k!)^2}(\bar{\phi} \phi)^{k} ~ .
\end{equation}
The minimal potential that supports stable $Q$-balls is
the one where $b_2$ and $b_3$ are nonzero and $b_k$'s with 
$k \ge 4$ vanish \cite{Col}.  Noncommutative generalization of 
this minimal setup is our primary interest.  Via the 
Weyl-Moyal correspondence, the replacement of products
between the fields in (\ref{comm}) with the $*$-products
\begin{equation}  
(f * g)(x)=\exp \left( \frac{i}{2} \theta^{\mu\nu}
 \frac{\partial}{\partial y^{\mu}}
 \frac{\partial}{\partial z^{\nu}} \right) 
 f(y)g(z)|_{y=z=x}
\end{equation}
should serve our purpose.  For each term in the potential $V$ 
beyond the quadratic order, however, there exist many 
inequivalent noncommutative generalizations; 
for example, depending on the ordering of fields,
the six possible interaction vertices (modulo cyclic
permutations)
\begin{eqnarray}
 \bar{\phi} * \bar{\phi} * \bar{\phi} * \phi * \phi * \phi ~ , ~
 \bar{\phi} * \bar{\phi} * \phi * \bar{\phi} * \phi * \phi ~ , ~
 \bar{\phi} * \bar{\phi} * \phi * \phi * \bar{\phi} * \phi 
 \nonumber\\
 \bar{\phi} * \phi * \bar{\phi} * \bar{\phi} * \phi * \phi ~ , ~
 \bar{\phi} * \phi * \bar{\phi} * \phi * \bar{\phi} * \phi ~ , ~
 \bar{\phi} * \phi * \phi * \bar{\phi} * \bar{\phi} * \phi 
\label{phi6}
\end{eqnarray}
are all different, while all of them reduce to 
$(\bar{\phi} \phi)^3$ in the commutative limit.  Among these
terms that have the global U(1) invariance, the term
\begin{equation}
\bar{\phi} * \phi * \bar{\phi} * \phi * \bar{\phi} * \phi
\end{equation}
can be singled out as the one that is also invariant
under the local U(1) invariance.  We also note that the 
potential term of the form 
$\bar{\phi} * \phi * \bar{\phi} * \phi$ was 
shown to be one-loop renormalizable in \cite{ABK}.  In this
paper, we choose to only consider the interaction terms
of the form
\begin{equation}
(\bar{\phi}\ast\phi)^{k} \equiv 
 (\bar{\phi}\ast\phi)\ast(\bar{\phi}\ast\phi)\ast
 \cdots\ast(\bar{\phi}\ast\phi) ~.
\label{pick}
\end{equation}
Implications of this choice will be addressed again later
in this paper.  We also note that we restrict our attention 
only to the spatially noncommutative field theories whose 
nonvanishing $\theta^{\mu \nu}$ components are $\theta^{x y}
 = - \theta^{yx} = \theta$.  When considering $Q$-balls,
this restriction significantly simplifies the analysis.
The noncommutative version of the action (\ref{comm})
is thus given by
\begin{equation}\label{act}
S=\int dt d^{2}x\left[\frac{1}{2}\partial_{\mu}\bar{\phi}\ast\partial^{\mu}\phi
-V(\bar{\phi},\phi)\right],
\end{equation}
where the potential is  
\begin{equation}\label{pot}
V(\bar{\phi},\phi)=\frac{1}{2}m^{2}\bar{\phi}\ast\phi+\sum_{k=2}^{N}
\frac{b_{k}}{(k!)^2}(\bar{\phi}\ast\phi)^{k} ~ .
\end{equation}

Conserved N\"{o}ther current due to the global U(1) symmetry is
\begin{equation}
j_{\mu}=-\frac{i}{2}[\bar{\phi}*\partial_{\mu}\phi-(\partial_{\mu}\bar{\phi})
*\phi], \label{noether}
\end{equation}
which yields the corresponding N\"{o}ther charge
\begin{eqnarray}
Q&=&\int d^{2}x j_{0}\nonumber\\
&=&-i\pi\theta\,{\rm
tr}[\bar{\phi} * \partial_{0} \phi-(\partial_{0}\bar{\phi})*\phi].
\label{cha}
\end{eqnarray}
When constructing the N\"{o}ther current (\ref{noether}), we 
utilize the property that 
the terms of the form (\ref{pick}) are invariant under the
local U(1) symmetry.  Inclusion of the other terms shown 
in (\ref{phi6}) in the action (\ref{act}) 
produces extra terms beyond the quadratic order in (\ref{noether}).
In the second line of (\ref{cha}), we use the one-dimensional simple
harmonic oscillator basis~\cite{GMS}, which is known to be convenient
for the construction of soliton solutions.   
The noncommuting spatial coordinates are related to the 
creation and annihilation operators via\footnote{Throughout
this paper, the coordinates $x$ can either mean commuting
coordinates or the operators defined on the noncommutative
space.}
\begin{equation}
a=\frac{1}{\sqrt{2\theta}}(x+iy),~~~
a^{+}=\frac{1}{\sqrt{2\theta}}(x-iy) ~~~ \rightarrow ~~~
 [ a, a^+ ] = 1~, 
\end{equation}
and
\begin{equation}
   \int d^2 x ~~~ \rightarrow ~~~ 2 \pi \theta ~ {\rm tr} ~ .
\end{equation}
The trace is taken with respect to this basis and the spatial 
derivatives inside the trace are understood as
\begin{eqnarray}
\partial_+ \phi &=& (\partial_{x} + i \partial_{y}) \phi
   = \sqrt{\frac{2}{\theta}} ~ [ a , \phi ] ~ , \nonumber \\ 
\partial_- \phi &=& (\partial_{x} - i \partial_{y})\phi
   =  \sqrt{\frac{2}{\theta}} ~ [ \phi , a^+ ] ~ .
\end{eqnarray}
In terms of the harmonic oscillator basis, the energy can be
written as
\begin{eqnarray}
E&=&\int d^{2}x\left\{\frac{1}{2}|\partial_{0}\phi|^{2}
+\frac{1}{2} |\partial_{i} \phi|^{2} 
+V( \bar{\phi}, \phi )\right\}\nonumber\\
&=&2\pi\theta\, {\rm tr}\left\{\frac{1}{2}|\partial_{0}\phi|^{2}
+ \frac{1}{2 \theta}[a, \phi] [\bar{\phi} , a^+ ] 
+ \frac{1}{2 \theta}[a^+ , \phi] [\bar{\phi} , a ]
+ V(\bar{\phi} , \phi )\right\} ~ .
\label{ener}
\end{eqnarray}
$Q$-balls are stable field configurations with the minimum 
energy for a fixed, conserved U(1) charge $Q$.  In the commutative
setup, the stationary form of the fields
\begin{eqnarray}\label{stat}
\phi=e^{i\omega t}|\phi|(x,y) ~~~ , ~~~
\bar{\phi}=e^{-i\omega t}|\phi|(x,y)
\end{eqnarray}
satisfies the minimum energy condition \cite{Col}.  The
time derivative terms in (\ref{act}) are quadratic and 
they are identical to those in the commutative case.  Furthermore,
the $*$-products that appear in the potential involve
only the spatial derivatives due to the restriction
to the spatial noncommutativity.  One can then
show that the same stationary ansatz (\ref{stat}) provides 
us with the minimum energy configuration for a given 
charge $Q$ in the noncommutative setup as well.    
For the stationary configurations, the charge (\ref{cha}) 
becomes
\begin{equation}
Q = \omega\int d^{2}x|\phi|^{2} =2\pi\theta ~ 
 \omega\, {\rm tr}|\phi|^{2}.
\end{equation}

Following \cite{GMS}, we analyze the system in the large
$\theta$ limit.  In this limit, the charge $Q$ scales
as $\theta$, which implies that the time derivative
term scales as $\theta$ just like the potential term.
Meanwhile, the spatial kinetic terms are of the order of
$\theta^0$ and can be neglected in the sense 
of \cite{GMS}.  
After neglecting the spatial derivative terms in
the energy expression (\ref{ener}), it becomes
\begin{eqnarray}
E&=&\int d^{2}x\left\{\frac{1}{2}|\partial_{0}\phi|^{2}
 +V ( |\phi | ) \right\}
\nonumber \\
&=&\frac{Q^{2}}{2 I}+
 2 \pi \theta ~ {\rm tr} ~ V ( |\phi |) ~ ,
\label{aene}
\end{eqnarray}
where
\begin{equation}
I=\int d^2x |\phi|^2 = 2 \pi \theta ~ {\rm tr} ~ |\phi|^{2} ~.
\end{equation}
Under the stationary ansatz (\ref{stat}), the potential
depends only on $|\phi| (x,y)$. 
Corresponding equation of motion in the infinite $\theta$-limit 
is given by
\begin{eqnarray}
\ddot{\phi}=
 -\frac{\phi}{|\phi|}\frac{dV}{d|\phi|} ~~~ \rightarrow ~~~
  \frac{dV}{d|\phi|} - \omega^2 | \phi | = 0 ~ ,
\label{eom}
\end{eqnarray}
where we use the stationary ansatz (\ref{stat}).   
Equation (\ref{eom}) is an algebraic equation involving $*$-products
solved in \cite{GMS}  
\begin{eqnarray}\label{eqeff}
\frac{d}{d|\phi|}V_{\rm eff}=0 ~ ,
\end{eqnarray}
where the effective potential $V_{\rm eff}$ is given by
$V_{\rm eff}=V-\frac{1}{2}\omega^{2}|\phi|^{2}$. 
Therefore, the general (radially symmetric) solutions
can be immediately written down as
\begin{equation}\label{gensol}
|\phi| = \sum_n a_n |n><n|,  \quad a_n \in \{\lambda_i\} ~ ,
\end{equation}
where  $\lambda_i$'s are the local extrema of $V_{\rm eff}$.
The corresponding energy and frequency $\omega$ are
determined as 
\begin{eqnarray}
E &=& \frac{Q^2}{2I} + 2 \pi \theta \sum_n V(a_n) ~ , \nonumber\\
\omega & = &  \frac{Q}{I} ~ , \label{whoa}
\end{eqnarray}
with $ I = 2\pi\theta \sum_n a_n^2$.
We note that $\lambda_i$'s are not extrema of $V$ but
of $V_{\rm eff}$, which in turn depend on $\omega$.
The $\omega$ expression in (\ref{whoa}) is thus 
implicit.  We also mention that one
has to go through the coherent basis analysis to obtain
the position space form of the solutions.  This procedure
is identical to the one given in \cite{GMS}.

Given the solutions (\ref{gensol}) satisfying the stationary
ansatz (\ref{stat}), the next step is to identify 
physical solutions that are classically and quantum mechanically 
stable.  We start from the classical stability
analysis to see if there exist any negative modes for the small
fluctuations around the classical solutions.
In the case of GMS solitons \cite{GMS}, the classical
stability of the soliton solutions requires that $\lambda_i$'s 
be local minimum points.  As we will demonstrate shortly, however, 
there are some exceptional cases where this feature changes for 
noncommutative $Q$-balls. 
Following the analysis of \cite{GMS}, a simplifying condition for 
the classical stability analysis is that we consider the 
large $\theta$ limit neglecting the spatial kinetic terms.   
It can also be shown that it is sufficient to consider only 
radially symmetric fluctuations. 
Under the variation $\delta | \phi |$ of the solution
(\ref{gensol}) for a fixed $Q$
\begin{equation}
\delta |\phi| = \sum_n \delta_n |n><n| ~ ,
\end{equation}
the energy changes as
\begin{eqnarray}
\delta E &=& \frac12 \int V''_{\rm eff} (\delta |\phi|)^2 
           + \frac{2 \omega^2}{I} 
	     \left( \int |\phi| \delta|\phi|\right)^2 + 
          {\cal O}(\phi^3) \nonumber\\
 & \equiv & \frac12 \sum_{m,n} \Delta_{mn} \delta_m \delta_n
  + {\cal O}(\phi^3) ~ ,
\end{eqnarray}
where
\begin{equation}
\Delta_{mn} = 2 \pi \theta V''_{\rm eff}(a_n) \delta_{mn} 
       + 4 (2 \pi \theta)^2\frac{Q^2}{I^3} a_m a_n ~ .
\end{equation}
For the moment, we assume that all $a_n$'s are either zero 
or the local minimum point $\lambda$.  The number of nonzero
$a_n$'s is denoted as $N$.  The U($\infty$) symmetry, which
is exact in the large $\theta$ limit, allows us to set
$a_n = \lambda$ for $n=0,\ldots,N-1$ and $a_n = 0$ for $n \ge N$, 
without loss of generality. 
The matrix $\Delta$ then becomes
\begin{equation}
\Delta = 4(2\pi\theta)^2\frac{Q^2 \lambda^2}{I^3}
         \pmatrix{ A^{(1)} & 0 \cr 
                   0 & A^{(0)} } ~ ,
\end{equation}
where $A^{(1)}$ is an $N \times N$ matrix 
($a, b = 0 , 1, \cdots , N-1$) 
\begin{equation}
A^{(1)}_{ab} = 1 + \alpha \delta_{ab} ~,
\end{equation}
with
\begin{equation}
\alpha = \frac{I^3}{8\pi\theta Q^2\lambda^2}V''_{\rm eff}(\lambda) ~ ,
\end{equation}
and $A^{(0)}$ is proportional to the identity matrix
($c, d = N, N+1 , \cdots $)
\begin{equation}
A^{(0)}_{cd}  = \frac{I^3}{8 \pi \theta Q^2 \lambda^2} V''_{\rm eff}(0)
                 ~ \delta_{cd} ~.
\end{equation}
In the present case the quantity $I$ is reduced to 
$I = 2\pi\theta N \lambda^2$.
As far as the $\phi = 0$ is a stable local minimum, the 
eigenvalues of the matrix $A^{(0)}$ are positive.  The $N$
eigenvalues of $A^{(1)}$ are $\alpha$ ($(N-1)$-times
degenerate) and $\alpha + N$.  Thus, as far as $\phi = 
\lambda$ is a stable local minimum, the solution
\begin{equation}
|\phi| = \sum_{n=0}^{N-1} \lambda |n><n|
\label{realthing}
\end{equation}
is classically stable.

We now consider the more general case when some of $a_n$'s are the local 
maximum point $\lambda'$ with the condition
$\lambda' < \lambda$.  Two numbers $N$ and $M$ represent 
the numbers of $\lambda$ and $\lambda'$ appearing in $a_n$'s, 
respectively. The nontrivial part of the matrix $\Delta$ is then
$(N+M) \times (N+M)$ submatrix $\Delta^{N+M}$ given by 
\begin{equation}
\Delta^{N+M} = 4 (2 \pi \theta)^2\frac{Q^2 \lambda^2}{I^3}
\pmatrix{ A^{(1)} & A^{(2)} \cr  A^{(2)} &  A^{(3)} } ,
\end{equation}
where $A^{(1)}$, $A^{(2)}$ and $A^{(3)}$ matrices are
given by ($a,b = 0, 1 , \cdots, N-1$ and $c,d = N , \cdots, N + M-1$)
\[ A^{(1)}_{ab} = 1 + \alpha \delta_{ab} ~~~ , ~~~
\]
\[ A^{(2)}_{ac} =  A^{(2)}_{ca} = \eta ~~~ , ~~~ 
\eta   = \frac{\lambda'}{\lambda} ~, \] 
\[ A^{(3)}_{cd} = \eta^2 + \gamma \delta_{cd} ~~~ , ~~~
 \gamma = \frac{I^3}{8 \pi \theta Q^2 \lambda^2}
               V''_{\rm eff}(\lambda') < 0  ~ , \]
and $ I = 2\pi \theta ( N \lambda^2 + M {\lambda'}^2 )$. 
We observe that the determinant of $\Delta^{N+M}$ is proportional
to $\alpha^{N-1} \gamma^{M-1}$; $\Delta^{N+M}$ has the negative 
eigenvalue $\gamma$ if $M>1$.  In other words, the solution 
is classically unstable when there are more than one local 
maximum points in $a_n$'s.  The case of $M=1$ is special and needs
careful treatment; the eigenvalue equation for the
$\Delta^{N+1}$ matrix reads 
\begin{equation}
 \det \left[
\pmatrix{ A^{(1)} & A^{(2)} \cr  A^{(2)} &  A^{(3)} } - \mu \right]
 = (\alpha - \mu)^{N-1} 
                [ \mu^2 - (\alpha + N + \gamma + \eta^2)\mu + 
                 (\alpha + N)\gamma + \eta^2\alpha] = 0 ~ .
\end{equation}
We find that if the condition 
\begin{equation} \label{gamma}
-\frac{\eta^2 \alpha}{\alpha + N} < \gamma < 0
\end{equation}
is satisfied, the eigenvalues of $\Delta^{N+1}$ are all 
positive.  Contrary to naive expectations, there 
exists some possible parameter region where the solution is 
classically stable even when one of $a_n$'s is at a local 
{\em maximum} point of $V_{\rm eff}$.  
We may call this solution an ``excited $Q$-ball." 
For the 
potential  of the 
form $V = a (\bar{\phi}* \phi ) + b ( \bar{\phi} * \phi )^2
 + c (\bar{\phi}* \phi )^3$ (renormalizable in the commutative 
setup in three dimensions), however, we emphasize that the
condition (\ref{gamma}) cannot be satisfied.  For the 
effective potential $V_{\rm eff}$
with sufficiently `flattened' shape near $\phi = \lambda^{\prime}$,
the condition can be met.  We further note that the energy
for this case is higher than that of the solutions (\ref{realthing})
as will be shown later.

Quantum mechanically, $Q$-balls might dissociate into perturbative
charged mesons.  To ensure the stability against this decay channel, 
we have to compare the $Q$-ball energy to the threshold energy of
the meson of the same charge at the tree level.  If the condition
\begin{equation}\label{emq}
E \le mQ
\end{equation}
holds, such decay is energetically disfavored.  
Since it is natural to expect that the solutions with 
each $\{a_n\}$ corresponding to a local minimum will
have the lower energy than other solutions, we first
demonstrate (\ref{emq}) in this situation.  We
assume, without loss of generality, that the number
of nonzero $a_n$'s is $N$ and they have the local 
minimum value $\lambda$.  The energy is then written as
\begin{equation}
E = 2 \pi \theta \left\{
  \frac{Q^2}{2(2\pi\theta)^2 N \lambda^2} + N V(\lambda) \right\}~. 
\label{qenergy}
\end{equation}
For a given $Q$, we have to choose the value of 
$N = N_{\rm min}$ such that 
the energy expression (\ref{qenergy}) is minimized.  
The appropriate value is  
\begin{equation}
N_{\rm min} = {\rm int} ~ \left[ \frac{Q}{2\pi\theta \lambda} 
 \frac{1}{\sqrt{2V(\lambda)}} \right] + N_1 ~ ,
\end{equation}
where ${\rm int} [x]$ is the largest integer not larger
than $x$ and $N_1$ is zero or one depending on the situation. 
The minimum energy for sufficiently large $N_{\rm min}$ is
thus given by 
\begin{equation}
E = Q \frac{ \sqrt{2V(\lambda)}}{\lambda} ~ ,
\label{qe}
\end{equation}
and the value of $\omega$ in this case is  
\begin{equation}
\omega = \frac{\sqrt{2V(\lambda)}}{\lambda} ~ .
\label{omega}
\end{equation}
The expressions (\ref{qe}) and (\ref{omega}) are identical 
to the expressions derived for the commutative $Q$-balls
when we identify $\lambda$ as the height of the $Q$-ball.
Moreover, the stability condition (\ref{emq}) translates
to $\omega < m$ using (\ref{qe}) and (\ref{omega}), and
this is equivalent to the condition
\begin{equation}
V(\lambda) < \frac12 ~  V''(0) \lambda^2 ~ ,
\label{qs}
\end{equation}
a result that is identical to the one in the commutative
case.  When the potential value at $\lambda$ is smaller
than the value of the quadratic part, $Q$-balls are stable
against decay into charged mesons.  In general, as far
as we neglect the spatial kinetic terms, 
the expressions (\ref{qe}) and (\ref{omega}) are valid for
any values of $Q$.  We note that there are two limits where the 
spatial kinetic terms can be neglected; the large noncommutative 
parameter $\theta$ limit and the large $Q$ limit (which is the
case of the large $Q$ thin-wall approximation limit in the 
commutative $Q$-ball setup).  Although they are different 
limits, the above calculations are valid for both cases.
In fact, the stability condition (\ref{emq}) can be generally
proved by the following line of arguments, which is valid even
when we include the spatial kinetic terms: 
From the equation of motion, we have
\begin{equation}\label{rel}
 |\phi | \ast \partial_i \partial_i |\phi | + 
 \omega^2 |\phi| \ast |\phi| = |\phi| \ast \frac{dV}{d|\phi|} ~ .
\end{equation}
Plugging (\ref{rel}) into the energy (\ref{ener}) leads to
\begin{eqnarray}
E  &=& mQ - mQ + \frac12 \omega Q + \int d^{2}x V \nonumber \\
   &=& mQ + 2 \pi \theta {\rm tr}\left\{
             \omega(\omega-m) + V - \frac12 |\phi|
  \frac{dV}{d|\phi|} \right\} ~ .
\label{whew}
\end{eqnarray}
We then use a H\"older-like inequality 
\begin{equation}
{\rm tr} |\phi|^2 {\rm tr} |\phi|^6 \ge ({\rm tr} |\phi|^4)^2 ~ ,
\end{equation}
which can be proved by diagonalizing $|\phi|$ and comparing both
sides term by term.  Following the arguments of \cite{KK}, the quantity
within in the curly bracket of (\ref{whew}) can be shown to
be negative, which implies the condition (\ref{emq}), 
at least for the range of $\omega$ given in \cite{KK}.

The solutions with one $a_n$ corresponding to the local maximum
point, the ``excited $Q$-balls", have higher energy 
than those considered above.  Quantum
mechanically, the former should decay into the latter.
To demonstrate this aspect, we repeat the same calculation as 
the one given from (\ref{qenergy}) to (\ref{omega}).
The value $N = N_{\rm min}$, which gives the minimum energy
for the ``excited $Q$-balls", turns out to be 
\begin{equation} \label{eN}
N_{\rm min} = {\rm int} \left[ \frac{Q}{2\pi\theta \lambda} 
    \frac{1}{\sqrt{2V(\lambda)}} 
           - \eta^2 \right] + N_1 ~ .
\end{equation}
The corresponding energy for the sufficiently large
$N_{\rm min}$ is
\begin{equation} \label{eqe}
E = \frac{Q}{\lambda} \sqrt{\frac{V(\lambda)}{2}} 
  + 2\pi\theta [ V(\lambda') - \eta^2 V(\lambda) ] ~ . 
\end{equation}
The energy of an ``excited $Q$-ball'' is larger
than that of a $Q$-ball by the second term in (\ref{eqe}). 
For the fixed charge $Q$, we can take the 
commutative limit where $\theta$ goes to zero while 
keeping the product $\theta N_{\rm min}$ fixed 
(see (\ref{eN})); we can still neglect the
spatial kinetic term as long as the charge $Q$
is large enough ($Q$-matter limit).  Upon taking
this limit the ratio,  $ - \eta^2 \alpha
 / ( ( \alpha + N )  \gamma ) $, goes like ${\cal O} (N^{-1})$ 
in (\ref{gamma}),
showing that the condition cannot be satisfied.
The classically stable ``excited $Q$-balls" do not exist 
in this limit.  The classical stability of the ``excited $Q$-balls"
depends on the interplay between the noncommutativity
and the existence of the conserved charge $Q$.

In the commutative field theory context, inclusion of the 
perturbative loop corrections in the quantum stability analysis
does not 
qualitatively change the arguments given in the above.  
To be specific, upon including 
the one-loop corrections, the relation (\ref{emq}) is replaced by 
$E_{Q}+\delta E_{Q}\le (m+\delta m)Q$ where $\delta E_{Q}$ and
$\delta m$ are one-loop contributions.  One can then consider 
the perturbative regime where the (renormalized) loop corrections
are smaller than the tree contributions.  In certain noncommutative
field theories, however, the UV/IR mixing appears \cite{MRS}; 
in noncommutative
$\phi^4$ theory in four dimensions, for example, the quadratic IR 
singularity present in the one-loop two-point 1PI amplitude drastically 
changes the low energy dispersion relations, casting doubts 
about the features observed at the classical level. 
In the present context with the global U(1)-respecting interaction 
terms (\ref{pick}) that also respect the local U(1) symmetry, 
the situation at the one-loop level is rather similar to that of 
commutative field theories.  

In (2+1)-dimensions, the one-loop two-point 1PI amplitude has the linear
UV divergence as can be seen by power counting, a potential
source of the UV/IR mixing.  The relevant interaction term
is the $\bar{\phi} * \phi * \bar{\phi} * \phi$ term whose 
interaction vertex in the momentum space can be read off 
from \cite{ABK}
\begin{equation}
 \frac{b_2}{4} \frac{1}{(2 \pi )^3}
 \int dp_1 dp_2 dp_3 dp_4 \delta ( p_1 + p_2 + p_3 + p_4 )
\label{vertex}
\end{equation}
\[   \times \cos \left( \frac{1}{2} 
 ( p_1 \theta p_2 + p_3 \theta p_4 ) \right)  
  \bar{\phi} (p_1 ) \phi (p_2 ) \bar{\phi} 
(p_3 ) \phi (p_4 ) ~ . \]
One immediately notes that the $\theta$-dependence in 
each possible one-loop two-point diagram drops out, demonstrating
the lack of the UV/IR mixing.  The commutative (renormalized)
perturbative treatment still applies in our context as well.
This analysis suggests that 
the qualitative features of $Q$-balls in consideration 
do not drastically change upon including the one-loop correction.
Had we included the term of the type 
$\bar{\phi} * \bar{\phi} * \phi * \phi$ in the 
action (\ref{act}), the UV/IR mixing does occur.  
In addition, the current expression (\ref{noether}) gets modified and
this can change the existence and properties of possible $Q$-ball 
solutions. At the higher loop level, the situation is more subtle.
The two-loop corrections to the 1PI two-point amplitudes 
include the ``sunset" diagram.  By power counting
the diagram has the logarithmic UV divergence in the commutative
setup and the $\theta$-dependence (determined by (\ref{vertex}))
appears not to cancel out, suggesting the possible logarithmic
UV/IR mixing.  The detailed study of the two-loop corrections is
thus clearly needed, but it is beyond the scope of this paper.

\section*{Acknowledgements}
We are grateful to Jin-Ho Cho and Jaemo Park for useful discussions.
This work is supported by No. 2000-1-11200-001-3 from the Basic Research
Program of the Korea Science $\&$ Engineering Foundation.

\end{document}